# How Can AI be Distributed in the Computing Continuum? Introducing the Neural Pub/Sub Paradigm

Lauri Lovén, Roberto Morabito, Abhishek Kumar,
Susanna Pirttikangas, Jukka Riekki, Sasu Tarkoma

*Abstract*—This paper proposes the neural publish/subscribe paradigm, a novel approach to orchestrating AI workflows in large-scale distributed AI systems in the computing continuum. Traditional centralized broker methodologies are increasingly struggling with managing the data surge resulting from the proliferation of 5G systems, connected devices, and ultra-reliable applications. Moreover, the advent of AI-powered applications, particularly those leveraging advanced neural network architectures, necessitates a new approach to orchestrate and schedule AI processes within the computing continuum. In response, the neural pub/sub paradigm aims to overcome these limitations by efficiently managing training, fine-tuning and inference workflows, improving distributed computation, facilitating dynamic resource allocation, and enhancing system resilience across the computing continuum. We explore this new paradigm through various design patterns, use cases, and discuss open research questions for further exploration.

*Index Terms*—Computing Continuum, Publish-Subscribe, AI Workflow Management, Resource Orchestration, 6G, AI-Native, Distributed Systems, Edge Computing

## I. INTRODUCTION

AS the number of connected devices and the amount of real-time data processing increases with the rise of 5G and beyond, traditional methods for managing the flows of information in large-scale distributed AI systems are increasingly insufficient. Existing approaches, predominantly reliant on centralized brokers, face challenges in scaling and meeting latency and privacy requirements of the device-edge-cloud computing continuum (also known as the computing continuum) [1]. Moreover, such traditional systems struggle to cope with fluctuating connectivity, heterogeneous and opportunistic compute, and distributed data inherent to the continuum. These challenges are further amplified by the rise of in-network AI, which refers to the use of AI to optimize network operations. While in-network AI has substantial potential to bolster these next-generation networks through machine learning operations (MLOps) and offer significant benefits to applications, it exacerbates the need for effective information flow management from data sources to subscribers within a large-scale, distributed AI system.

Addressing these limitations, we introduce a new solution: the neural publish/subscribe paradigm. Leveraging the decentralization aspect of AI, it aims to effectively orchestrate machine learning data flows within the computing continuum. The neural publish/subscribe paradigm, our proposed solution, integrates AI into the communication fabric of the computing continuum. It builds upon the publish/subscribe model, which decouples communicating endpoints in space, time, and synchronization, and enables many-to-many deep learning-based information dissemination and distributed inference and learning. This unique approach paves the way for efficient information flow management in the computing continuum and can support diverse machine learning models for textual, aural, and visual content. The Neural Pub/Sub paradigm thus has the potential to substantially enhance in-network AI and MLOps capabilities.

In this position paper, we delve into the neural publish/subscribe paradigm, expounding its underlying principles, potential advantages, and use cases such as 5G/6G Mobile Network, Metaverse and foundation models.

## II. THE NEED FOR A NEW PARADIGM

AI is increasingly integrated into the 6G architecture [2] and wireless networks and communications [3]. Consequently, reevaluating machine learning (ML) distribution and workflow management is vital for the next generation of networks and wireless systems. This section outlines the requirements for ML workflow management in the computing continuum, briefly looks at state-of-the-art and its shortcomings in relation to those requirements. Finally, this section proceeds to describing the publish/subscribe paradigm, which has potential for fulfilling the ML workflow management requirements in the computing continuum.

### A. Requirements

The computing continuum represents a distributed computing paradigm that spans from centralized cloud data centers to edge devices close to data sources. It poses a novel set of challenges that mandate a rethinking of the requirements of ML workflow management systems deployment.

**Resource efficiency and management.** ML workflow management systems for optimal resource usage and effective resource management must account for stringent computational constraints of edge devices across the continuum into account [4].

L. Lovén (lauri.loven@oulu.fi), A. Kumar (abhishek.kumar@oulu.fi), S. Pirttikangas (susanna.pirttikangas@oulu.fi) and J. Riekki (jukka.riekki@oulu.fi) are with the Center for Ubiquitous Computing, University of Oulu, Finland.
R. Morabito (roberto.morabito@helsinki.fi) and S. Tarkoma (sasu.tarkoma@helsinki.fi) are with the Department of Computer Science, University of Helsinki, Finland.

**Latency and bandwidth.** Latency and bandwidth emerge as pivotal factors. Depending on the use case, low latency may be prioritized, necessitating more localized processing at the edge. Bandwidth constraints could also limit data transmission back to the cloud, underscoring the need for more efficient data handling and potentially local data processing capabilities.

**Model adaptivity.** The nature of the ML models themselves may need to adapt to this environment. Due to the resource constraints at the edge, the deployment of smaller and less computationally intensive models may be necessary. Thus, the system should support model compression, quantization, or other techniques for reducing model size and complexity [5], or support distributing the models across the computing continuum [4].

**Data privacy and security.** Edge processing can enhance privacy as data may never leave the local device. However, given the potential vulnerabilities of edge devices and the requirement of data transmission over potentially insecure networks, distinct security measures need to be accounted [6].

**Model synchronization and updates** pose additional challenges. As models might be trained or updated in different parts of the continuum, effective management and synchronization of these updates are required. This might necessitate techniques such as federated learning [7].

**Robustness and fault tolerance.** The system must be designed for robustness and fault tolerance, given that edge devices may have less reliable connectivity and might be more prone to failures [8].

**Heterogeneity.** Computing continuum comprises of a wide variety of hardware, software, and network configurations [1]. The ML workflow management system must be able to handle such a heterogeneous ecosystem, ensuring interoperability across devices.

*B. State-of-the-art*

State-of-the-art systems like Kubeflow and MLflow manage ML workflows by leveraging deterministic orchestration, where each step's execution is explicitly defined and known in advance. While this approach offers a high degree of control and predictability, it may also lead to certain challenges. One key issue stems from its non-dynamic nature: these systems may struggle to adapt swiftly to real-time changes in the data or the environment, which is a common requirement in today's ever-evolving data landscapes, and especially present in the highly dynamic computing continuum. Moreover, the predetermined sequence of operations can restrict the capacity to parallelize tasks that do not have inter-dependencies, potentially leading to inefficiencies in resource utilization. Further, the rigid structure could make it difficult to integrate new steps or modify existing ones in response to evolving requirements, which could impact the agility and flexibility of the ML development process. These potential limitations underscore the need for more dynamic and responsive orchestration mechanisms in the design of future ML workflow management systems.

For example, Argo Workflows[1], an integral part of Kubeflow Pipelines, manages and executes machine learning (ML) workflows in a Kubernetes-native environment. This workflow engine orchestrates the execution of each containerized step in the pipeline, ensuring that they run in a predefined sequence to form a directed acyclic graph. However, Argo's deterministic orchestration model, where each step is explicitly defined and known apriori, might limit flexibility in highly dynamic edge environments. For example, in a scenario where immediate response to real-time data is required, Argo's deterministic model may not allow for fast enough adjustment. Yet edge devices, due to their proximity to data sources, are well-suited to handle dynamic, real-time ML tasks. Therefore, a more event-driven, flexible orchestration could be beneficial. Also, the reliance on direct API calls for component communication might pose latency and bandwidth issues, especially considering the potential network constraints (e.g., intermittent or fluctuating connections) in edge environments.

Finally, Kubeflow depends on Kubernetes, a resource-intensive and centralized container orchestration framework. Even with edge extensions, Kubernetes doesn't easily run on resource-poor edge nodes, discouraging its use in heterogeneous computing continuum environments [1], [9].

*C. Publish/Subscribe*

The Publish/Subscribe (Pub/Sub) paradigm could potentially infuse more dynamism into ML workflow configurations, particularly with respect to reacting to real-time data or changes. Unlike traditional control flows that are deterministic and pre-defined, the Pub/Sub paradigm allows for more reactive and event-driven processes [10]. In this model, various stages of the ML pipeline could act as subscribers that listen for specific messages or data updates published by other stages (publishers). When these messages or updates are published, the corresponding subscribers would react and process the new data or adjust their behavior accordingly [11]. This could enhance the flexibility of the ML workflow, enabling it to adapt to real-time changes and potentially improving the robustness and responsiveness of the system [12]. Furthermore, the decoupling of data producers and consumers inherent in the Pub/Sub paradigm can lead to better scalability and fault-tolerance, as the failure of a single component doesn't directly impact others [10].

Fig. 1 depicts an AI workflow implemented with a traditional broker-based pub/sub architecture. AI processing is executed on the client-side at subscribing modules, which can publish their inference results, facilitating distributed patterns across the pub/sub infrastructure. However, this model encounters limitations since computation occurs in those subscribing clients, and optimizing this processing requires developing a cooperative framework for the clients. Moreover, subscriptions for model training, updating or inference remain underexplored in the literature.

To address the limitations of the existing state-of-the-art systems, we propose the Neural Pub/Sub paradigm for weaving

---

[1]https://github.com/argoproj/argo-workflows

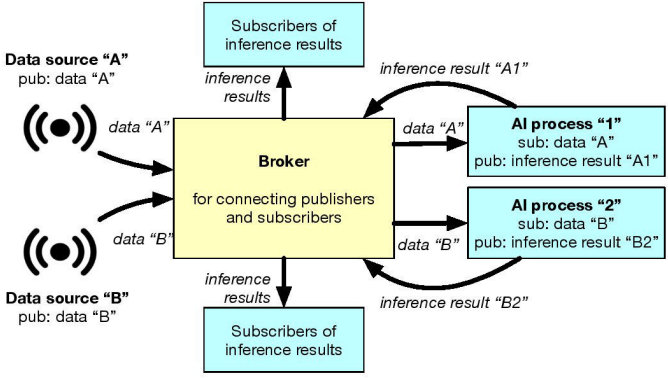

Fig. 1: Traditional broker-based pub/sub AI workflow. Processing occurs at client-side subscriber modules, which publish inference results. This design faces challenges, especially since optimization requires cooperative frameworks, and aspects like model training and updating are less explored.

AI into the communication substrate and within the device-edge-cloud computing continuum. Neural Pub/Sub addresses the limitations by combining the observer and the distributed broker models in a seamless manner, enabling system-wide placement and optimization of data, learning, and inference.

## III. NEURAL PUB/SUB

To address the requirements for AI workflow management in the computing continuum, we propose the Neural Pub/Sub architecture, a novel design that draws inspiration from distributed publish/subscribe models. It enhances the traditional models by raising learning and inference as first-class citizens within the system.

### A. Architecture and Testbed Deployment

Our proposed Neural Pub/Sub architecture brings a novel perspective to the publish/subscribe systems, offering the ability to subscribe to inferences and decompose machine learning operations into distributed pipelines. This innovative approach follows the principle of balanced upstream inference, aiming to execute as much computation as possible as close to the data source, while taking into account the resource constraints and system state. This strategy reduces the volume of data to be transferred and processed, enhancing the system's efficiency and scalability.

In Neural Pub/Sub, pipeline formation leverages the principle of decomposition, breaking down the AI process into smaller, simpler operations that can be performed independently and in parallel. This strategy distributes the computation across multiple devices, thus improving performance and scalability.

The architecture encompasses four key components, as illustrated in Fig. 2: the Publisher, the Neural Pub/Sub broker, the Execution Units, and the Subscriber. Each possesses a distinct functionality stack, underpinned by either 5G-based or semantic-based communication and orchestration layers (the latter relevant for the 6G transition).

In this ecosystem, the Broker plays a pivotal role in managing AI models discovery across the network. Its stack features an AI Models Discovery, an AI Models Splitter and Reasoner, and an AI pipelines scheduler. The Splitter and Reasoner component is responsible for instantiating and managing pub/sub AI inference pipelines in the distributed environment. This is achieved with the support of the AI Model Discovery, which maintains an overview of the AI Services available across the different nodes. These pipelines connect the publishers and subscribers of data and data-based inference, enabling a many-to-many form of communication.

The Publisher's stack, which includes the Publisher Capabilities Monitoring and Data layer, generates the information to be processed and sends it out to the system.

At the Subscriber's end (which receives the processed data and uses it for its own purposes), the stack includes a Subscriber Capabilities Monitoring layer and an AI Models Repository. These are flanked by two peer services: the subscriber's application (e.g., text, visual, aural, telemetry) that consumes the received data, and the Local AI Services, which can possibly apply additional AI-based processing to the subscriber's application.

Finally, the Execution Units are computational resources that actually perform the tasks assigned by the Neural Pub/Sub broker. It is the component in the system that carries out the work of running the AI models on the data. Execution Units could be located anywhere in the network, from edge devices to central servers, and their role is to execute the AI/ML computations as close to the data source as possible to maximize efficiency.

It is possible in some systems for the same physical component to act as both an Execution Unit and a Subscriber, but conceptually, they are separate roles in the Neural Pub/Sub system. An Execution Unit is a computational resource that performs the AI/ML tasks. It does the work of running the models on the data and produces the output or "AI inference results". The Subscriber, on the other hand, is the entity that consumes the output of the Execution Units. It receives the processed data and uses it for its own purposes. If the same physical component acts as both an Execution Unit and a Subscriber, it means that it's not only running the AI/ML tasks but also using the results of those tasks for some further purpose. For example, in an AR/VR application, a player's device might act as both an Execution Unit (processing sensor data to interpret the player's movements) and a Subscriber (using the interpreted movements to update the game state). However, it is crucial to keep in mind that even if the same physical component is performing both roles, they are still conceptually distinct roles within the Neural Pub/Sub system.

Orchestration, which involves gathering resource information and implementing subscribed inference pipelines based on various metrics (such as CPU, GPU, memory, AI accelerator details, configuration, and network information), plays a crucial role in the Neural Pub/Sub architecture. The system is expected to have resources across the edge-cloud continuum that can be activated for pipeline execution when necessary.

As the Neural Pub/Sub system evolves towards a 6G-oriented paradigm, the system will integrate semantic models into its architecture. This evolution implies that the semantic model-based communication and semantic-based orchestration will gradually replace the existing 5G communication and orchestration layers. The stacks of the subscriber, Neural Pub/Sub broker, and publisher remain consistent, but each now integrates a "Semantic Models" capability, thus enhancing their proficiency in communication and orchestration.

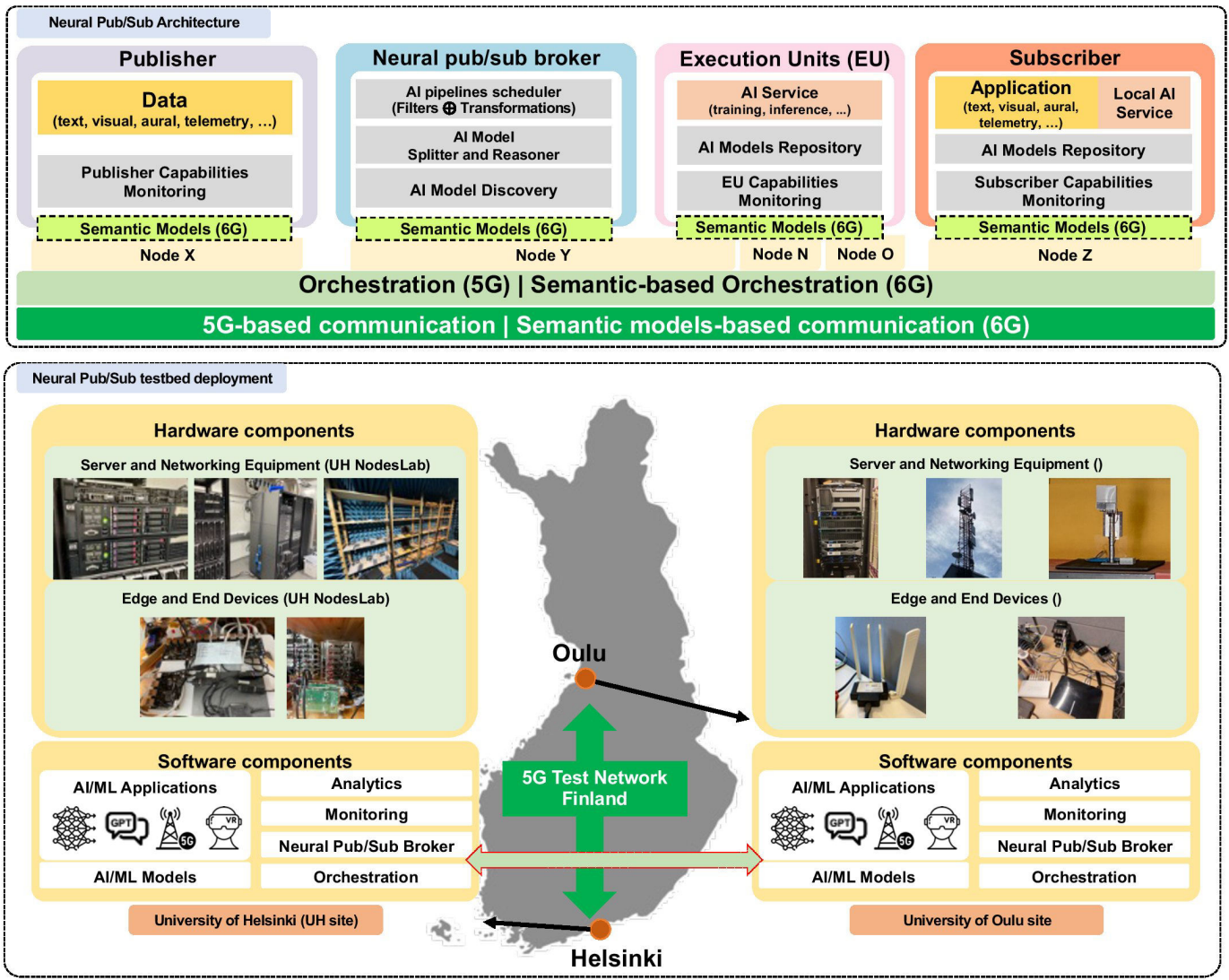

Fig. 2: This figure illustrates the three key components of the Neural Pub/Sub system (Publisher, Broker, and Subscriber), as well as the testbed development plan for the implementation of our solution. Each component features its own functionality stack, underpinned by 5G-based (or semantic-based for 6G transition) communication and orchestration layers. The Broker's central role in AI model management is evident, while the Publisher's role includes AI services provision and application enablement.

This architecture allows for many-to-many semantic compression of data, model split and partitioning, inference-based filtering and privacy protection, and the use of prediction as a basic primitive. It can support a wide range of AI models, including those for textual, aural, and visual content, thus paving the way for the future of AI workflow management in the computing continuum.

In the context of furthering the practical application and real-world validation of the Neural Pub/Sub architecture, a testbed has been developed in collaboration between two sites: the University of Helsinki's NodesLab and the University of Oulu's Center for Ubiquitous Computing. The two sites are connected through the 5G Test Network Finland (5GTNF), a state-of-the-art multisite test environment that provides a realistic telecom technology setting based on real vertical system and service needs and requirements. The 5GTNF supports research and large-scale field trials on 5G and beyond technologies, thus serving as a robust platform for our experimentation[2]. Within this testbed, we are actively exploring various applications setup, as discussed in detail in Section IV. The two sites' implementations are orchestrated seamlessly, linking the two Neural Pub/Sub brokers belonging to the two local administrative domains. This interconnection ensures coherent data flow and efficient processing across geographically dispersed locations, reflecting the potential of Neural Pub/Sub in handling complex multi-domain scenarios. The Neural Pub/Sub preliminary implementation, coupled with the advanced infrastructure of the 5GTNF, not only emphasizes the innovation of our approach but also substantiates its practical feasibility and adaptability in modern telecommunication and applications landscapes.

[2]More details can be found at https://5gtnf.fi/ .

*B. Functionality*

The Neural Pub/Sub paradigm, designed with the computing continuum in mind, addresses the requirements for efficient AI workflow management (see Section II-A) as follows:

**Resource efficiency and management**: Neural Pub/Sub extends the publish-subscribe paradigm not just to data but also to model inference and training, enabling efficient dissemination of information and computation across the continuum. Rather than requiring all nodes to process all data, nodes subscribe only to the inference or training they require which ensures optimal resource usage and effective management across the continuum.

**Latency and bandwidth**: Neural Pub/Sub introduces a funnel task, which combines multiple publications into a single emitted publication. This aids in reducing network latency and bandwidth usage, streamlining information flows.

**Model adaptivity**: Neural Pub/Sub supports many-to-many semantic compression of data, allowing the splitting and partitioning of ML models, a key aspect of model adaptivity in resource-constrained environments.

**Data privacy and security**: Inference-based filtering in Neural Pub/Sub can enhance privacy protection by ensuring that only necessary data is disseminated to subscribers. This feature aligns with the unique privacy considerations in the computing continuum discussed in Section II-A.

**Model synchronization and updates**: Neural Pub/Sub supports a wide range of ML models, maintained in a model repository. This aids in model synchronization across the continuum. The model splitter and reasoner components manage and instantiate pub/sub inference pipelines, allowing for effective synchronization of model updates.

**Robustness and fault tolerance**: Pub/Sub approach offers robustness and fault tolerance by design. If a publisher, subscriber, or broker fails, the overall system can still continue to operate.

**Heterogeneity**: By supporting a wide range of ML models for various types of content, Neural Pub/Sub ensures interoperability across devices with different hardware, software, and network configurations. Using *prediction* as a basic primitive aids in maintaining continuity in ML workflows, even amidst device heterogeneity and potential failures.

*C. Design Patterns*

The Neural Pub/Sub system utilizes two key design patterns to support distributed data processing: mapping and funnelling (Fig. 3). These patterns can be used to implement micro-batching style operations.

The mapping pattern is an elementary element that takes in a publication and applies a function (F) to it. This function can result in a transformed version of the original publication, or the original publication itself. This pattern is useful for local and distributed data processing, and can be used to implement simple data transformation and processing tasks.

The funnel pattern builds on the mapping pattern by subscribing to one or more publications and applying a function (F) to a combination of the received publications. This function can determine the order and timing of the processing. The funnel pattern results in a single emitted publication, which is based on the output of function F.

Neural Pub/Sub systems can utilize these two patterns to execute complex data processing tasks at very different levels of granularity: one at the AI model level (e.g., distributed AI training) and one at the AI-powered application level (e.g., distributed AI pipeline). The use of such patterns allows for the orchestration of data flows in the continuum. Following the scheme of Tirana et al. [13], we can segment an AI model into layers or an AI-powered application into sub-components. Each segmented layer or component subscribes to model data for a specific sub-component and publishes data for the next one. This means that the components can be relocated within the environment, supporting dynamic placement and parallel data processing. Mapping and funnel patterns enable the transformation of a complex task into a set of simple tasks that can be performed at or near the data site (publisher), reducing the need for data movement as required in centralized systems. If the task indeed needs to be performed by a third-party broker (i.e., lack of sufficient resources at the publisher to finish the task), privacy protection can be ensured by performing part of the training or application at the publisher and the rest at the broker. For example, in Fig. 3 (right), the AI inference application is split into $k$ sub-components, and $1^{st}$ and $k^{th}$ sub-components are always at the publisher.

IV. USE CASES

The potential of Neural Pub/Sub paradigm for handling and optimizing complex data flows and AI operations can be envisioned across diverse domains such as 5G/6G mobile networks, AR/VR/Metaverse, and applications utilizing foundation models like Large Language Models (LLMs).

**5G/6G mobile networks.** 5G/6G networks are key use cases for the Neural Pub/Sub paradigm. With the growing number of connected devices, sensors, and actuators in these networks, there is an escalating need for efficient and effective management of data flows and AI-based resource orchestration. Neural Pub/Sub addresses this need by providing a framework for monitoring RAN and Core Network KPIs and positioning AI processes close to the data sources, even while satisfying the needs of constrained devices such as base stations.

A defining feature of Neural Pub/Sub is its capability to monitor and update models in real-time through the use of event-based communication. This is particularly significant in the context of MLOps, where efficient monitoring and updating of models is vital for ensuring the performance and accuracy of AI-based systems. By offering a basic mechanism for monitoring inference processes and propagating model updates, Neural Pub/Sub can ensure efficient processing of network and application information flows, and appropriate distribution of AI, adhering to the principle of maximal upstream inference.

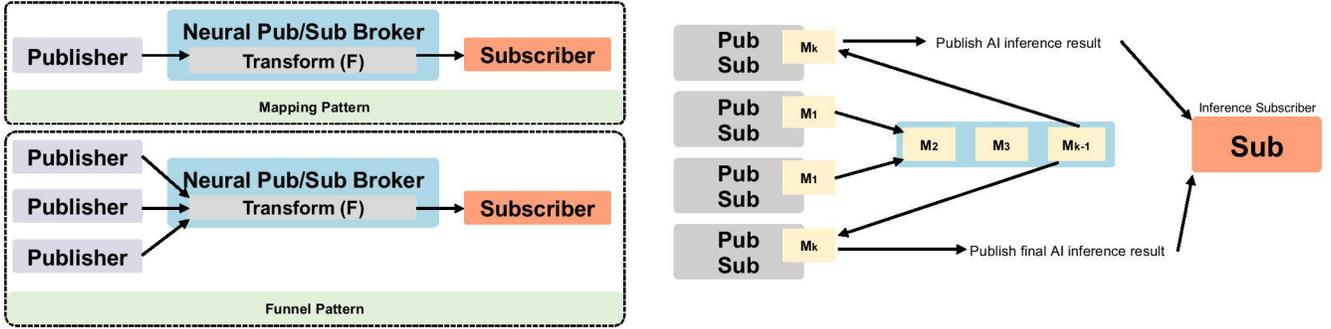

Fig. 3: On the left, illustration of the two key design patterns utilized in the Neural Pub/Sub paradigm. The mapping pattern involves taking a publication and applying a function to it, either transforming the original publication or leaving it unchanged. The funnel pattern builds upon the mapping pattern, combining one or more publications and applying a function to the received publications. On the right, an example of design patterns application in the case of distributed AI inference orchestrated by the Neural Pub/Sub paradigm.

Specifically, it can be instrumental within the Network Data Analytics Function (NWDAF) and Open-RAN architectures, which are fundamental to these infrastructures. In NWDAF (Fig. 4a), Neural Pub/Sub manages and disseminates the analytic models used for monitoring network Key Performance Indicators (KPIs). By placing AI processes near data sources like base stations, latency is reduced, allowing for real-time processing. The principle of maximal upstream inference lightens the load on the central server, optimizing resource allocation and boosting network efficiency. Within Open-RAN (Fig. 4b), Neural Pub/Sub optimizes the distribution of xApps and rApps in a multi-vendor environment. By dynamically deploying models close to the edge, it ensures swift, efficient, and resilient network operations.

**AR/VR/Metaverse.** Neural Pub/Sub can be used to link sensors and their AI/ML functionalities in a distributed environment, enabling efficient data sharing of AR/VR interaction and world data. It is vital in the Metaverse context, where the efficient placement and distribution of sensing and inference pipelines are crucial for ensuring low latency and real-time interaction.

In a distributed multiplayer game, for example, each player's device (Publisher) generates a stream of sensor data. The Neural Pub/Sub system breaks down necessary AI operations like noise filtering, gesture classification, action mapping, and game state computation into a pipeline distributed across edge servers (Execution Units). This approach ensures real-time updates and minimal latency, enhancing the immersive experience for the users. The many-to-many communication facilitated by Neural Pub/Sub significantly reduces overall bandwidth usage compared to centralized solutions. It enables semantic compression of data streams and optimizes the distribution of data and inference across the network (demonstrated in Fig. 5a).

**Foundation models.** Foundation models such as LLMs are quickly advancing to the edges of the network. They can handle a wide range of tasks related to the generation of modalities such as text, images, and video, including language translation, summarization, and question answering. The deployment and/or use of foundation models in the computing continuum may provide several benefits over cloud-based deployment, such as latency, bandwidth, or privacy [14].

For tasks like translation or summarization in a multi-device environment, Neural Pub/Sub facilitates the creation of an optimized pipeline. Components of this pipeline, such as initial language detection, text segmentation, translation, and post-processing, may run on different devices (Execution Units), yielding real-time results with reduced latency and network traffic. Subscribers to this pipeline can receive translation services in real-time, while the system's scalability can handle increased demand for these services, contrasting with the limitations of centralized solutions. Fig. 5b showcases the operation of the Neural Pub/Sub system with foundation models for Natural Language Processing (NLP).

The utilization of Neural Pub/Sub for foundation models allows for the efficient distribution and processing of large amounts of data, which is crucial for fine-tuning these models. Additionally, the ability to subscribe to foundation model inference, with the pipeline dynamically optimized by the platform, simplifies the training and deployment of these models in a distributed environment. This can lead to significant improvements in performance, scalability, and cost-effectiveness.

## V. OPEN RESEARCH QUESTIONS

We have identified four main categories of challenges and corresponding research questions surrounding the implementation and optimization of the Neural Pub/Sub system (also highlighted in Table I): 1) system design and optimization, 2) model design and optimization, 3) distributed learning and inference 4) real-world implementation.

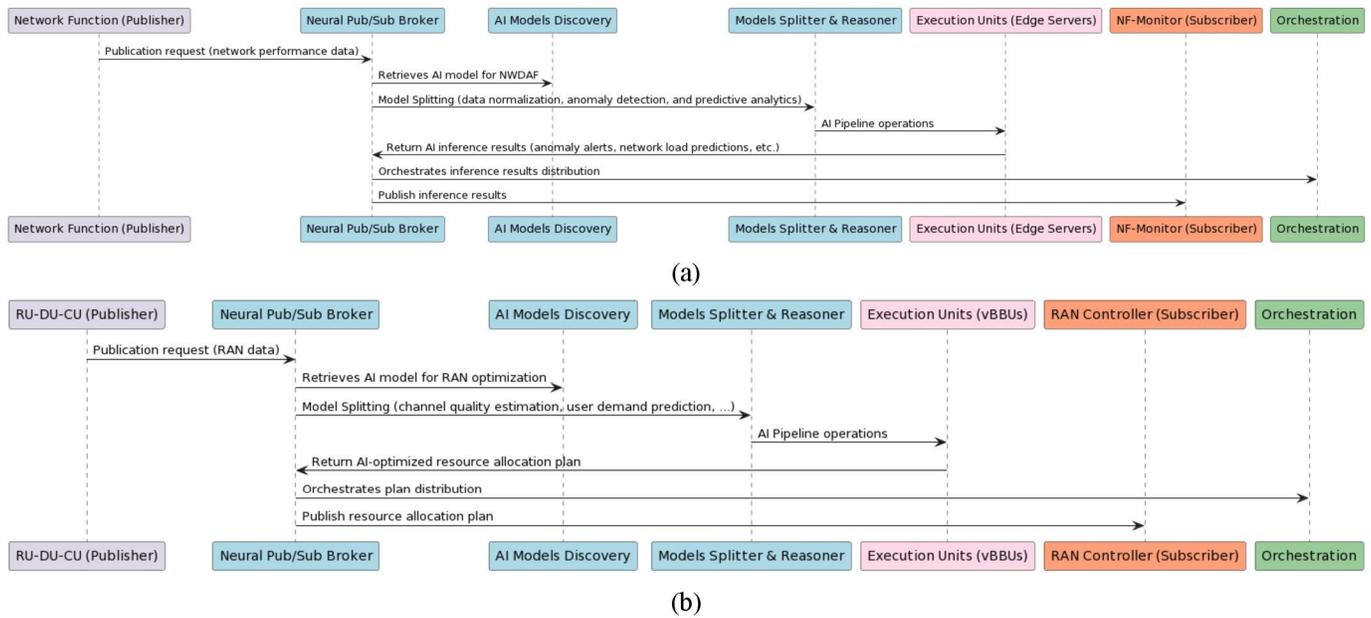

Fig. 4: (a) NWDAF: This diagram illustrates the implementation of Neural Pub/Sub within 5G/6G networks via NWDAF. Network Functions act as Publishers, broadcasting network performance data, while AI processes close to the data sources, or Execution Units, perform operations in a pipeline created by the Broker. The results are published to the Subscriber (NF-Monitor) for effective network management actions. (b) Open-RAN: The figure shows the application of the Neural Pub/Sub system in Open-RAN architecture. The RU-DU-CU stack generates real-time RAN data as the Publisher. The Broker creates a pipeline of operations executed by Execution Units, such as virtualized baseband units in vRAN. The optimized data is then published to the RAN Controller, enabling efficient network performance adjustments.

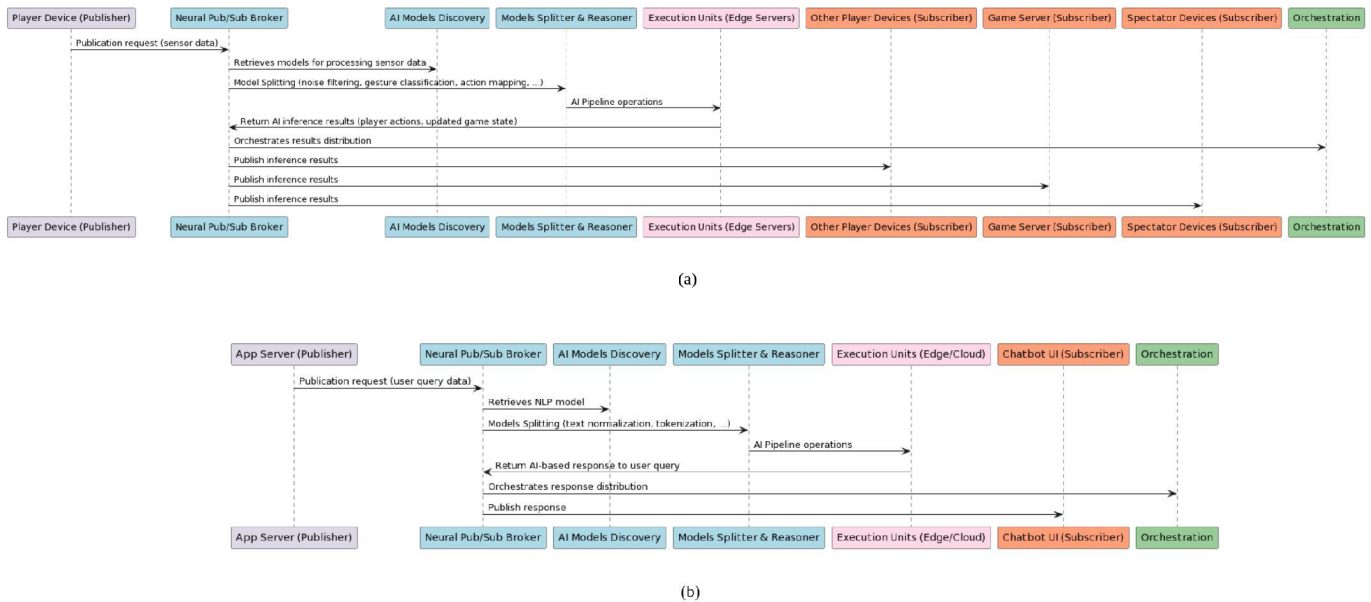

Fig. 5: (a) AR/VR/Metaverse: This diagram demonstrates the application of the Neural Pub/Sub system in an AR/VR/Metaverse scenario. Player devices (Publishers) generate a stream of sensor data which is processed through a pipeline of operations retrieved from the AI Models Repository. The pipeline operations are carried out by Execution Units, typically edge servers due to the real-time nature of the application, and the inference results are published to Subscribers, such as other player devices or a central game server. (b) Foundation models: The diagram portrays the use of the Neural Pub/Sub paradigm with foundation models for NLP. It shows how raw user query data from an App Server (Publisher) is processed using a decomposed pipeline of operations retrieved from the AI Models Repository. The operations are distributed across various Execution Units including edge servers and cloud resources, and the generated responses are published to the Chatbot UI (Subscriber).

8TABLE I: Challenges and open research questions.

| Challenge | Research Questions |
| --- | --- |
| System design and optimization | How can we design algorithms that can efficiently split and place models within the computing continuum? |
| | What are the trade-offs between different partitioning and distribution strategies in terms of computation, communication, and privacy? |
| | How can we design and implement efficient funnel mechanisms for combining multiple publications across the computing continuum? |
| | How can we handle conflicts and inconsistencies between multiple publications? |
| | How can we ensure the robustness and resilience of the Neural Pub/Sub distributed broker against network failures and unexpected changes in the computational environment? |
| Model design and optimization | Are models used for inference suitable for communication optimization? |
| | How can we adapt models for use in a distributed system, and what are the trade-offs involved? |
| | Are split and transfer learning viable solutions for Neural Pub/Sub networks, and if so, how can we implement it? |
| | How can we maintain model interpretability and transparency while optimizing communication in the Neural Pub/Sub system? |
| Distributed learning and inference | How can we effectively synchronize and coordinate distributed learning and inference in a Neural Pub/Sub system to optimize model performance and communication efficiency? |
| | What are the techniques to ensure security and privacy during distributed learning and inference, especially considering cross-boundary data exchange? |
| | How can we handle heterogeneous devices and various data formats in Neural Pub/Sub systems, ensuring seamless and interoperable learning and inference? |
| Empirical Results and Integration | What is the minimum viable product that can be developed for the Neural Pub/Sub system to obtain meaningful empirical results for testing and evaluation? |
| | How can we integrate Neural Pub/Sub into existing networks and systems, such as RAN and CN? |
| | What are the key metrics for evaluating the performance of Neural Pub/Sub in real-world systems, and how can we measure them? |

## A. Design and optimization

One of the key challenges in Neural Pub/Sub is designing algorithms that can optimally split and place AI models and processes in a distributed environment. This involves finding the best trade-off between computation and communication costs, as well as developing methods for efficiently distributing and updating these partitions, ensuring that the models are placed in the right location for maximum performance. However, determining the optimal way to split models and place them in the computing continuum is complex. Factors such as the varying computation and memory requirements of different models, the availability of resources in various locations, and the need to consider network latency and geography in determining optimal placement contribute to this complexity. Additionally, the optimal placement of models may change over time as the model's requirements or available resources change. This necessitates a dynamic approach that can adapt to shifting conditions, posing a significant challenge in itself.

Moreover, the funnel mechanism is a key feature of Neural Pub/Sub, allowing for the combination of multiple publications into a single emitted publication. Many open questions related to how this mechanism functions in practice persist, such as how to efficiently combine different publications and how to resolve conflicts between them. Determining the appropriate order and timing of processing for the publications presents a primary challenge. The order and timing can significantly impact the accuracy and effectiveness of the combined publication. Ensuring that the funnel mechanism scales well to handle a large number of publications while maintaining low latency and high throughput is another challenge. This requires careful consideration of the algorithms used to combine publications and the data structures used for storing and processing them. Additionally, designing a funnel mechanism that can manage publications of different formats and structures can be arduous, necessitating the creation of robust and flexible data processing algorithms capable of handling a wide array of data types and formats.

Lastly, privacy and security present a further challenge. Funneling data from various sources and combining them into a single publication may raise privacy concerns, and the system must provide means to address these issues.

## B. Model design and optimization

Implementing Neural Pub/Sub requires determining which models are suitable for inference and communication optimization. This includes understanding the properties of different models and how they can be adapted for use in a distributed system. Specifically, the key challenge is determining the appropriate level of granularity for the models. If the models are too granular, they may not capture the complexity of the data and may not be able to make accurate predictions. On the other hand, if the models are too coarse, they may be too computationally expensive to run in a distributed environment.

Another challenge is determining the appropriate type of models to use for different types of data. Different modalities may require different types of models to be processed efficiently. For example, a model that is optimized for text data may not be suitable for processing audio data.

Balancing the trade-off between model accuracy and communication efficiency remains a key issue. In general, more complex models may provide higher accuracy but also require more communication resources.

## C. Distributed learning and inference

One of the main challenges in distributed learning of models in Neural Pub/Sub is the difficulty in coordinating and synchronizing the updates to the models across the distributed system. This includes issues such as communication overhead, data privacy and security, and handling data inconsistencies and outliers. Additionally, traditional backpropagation algorithms may not be well-suited for distributed learning due to their high computational and communication costs. Understanding on how to effectively design split learning solutions is largely missing that can overcome these challenges in Neural Pub/Sub.

Coordinating and managing the distributed inference process across multiple nodes in the system is required to ensure that the necessary data and models are available at each node, and the results of the inference process are properly synchronized and shared across the system. Additionally, there may be challenges with communication and data transfer between nodes, particularly in terms of latency and bandwidth constraints. Ensuring that the distributed inference process is efficient and accurate, while also addressing these communication and coordination challenges, is a complex task that requires further research.

## D. Implementation of Neural Pub/Sub in real-world systems

Building a minimum viable product (MVP) of the Neural Pub/Sub requires determining the minimal set of features and functionality that are required to demonstrate the feasibility and potential value of the system. This involves identifying the key components and capabilities that are necessary for the system to perform its intended functions and meet the needs of its target users. Additionally, building an MVP requires careful planning and design to ensure that the system is both technically feasible and economically viable. This may involve making trade-offs between functionality and cost, and balancing the need for robustness and scalability with the need for rapid development and deployment. Another challenge is to ensure the system performs well under different use cases, scenarios and environments. The further development of the previously described distributed testbed is a crucial step towards achieving the goal of determining the minimal set of features and functionality needed to demonstrate the feasibility and potential value of the system. Especially in the context of integrating the Neural Pub/Sub paradigm into existing mobile network technologies such as RAN and Core Network, a deeper understanding of the requirements and constraints of the existing systems' architecture, communication protocols, and interfaces is required. This understanding is essential to adapting the Neural Pub/Sub system to work within those constraints. The adaptation of the Neural Pub/Sub system to work within these constraints is being explored, with significant testing and validation being essential to ensure that the integration does not negatively impact performance, functionality, security, or compliance with relevant standards and regulations. The testbed development and ongoing research serve as vital steps towards the realization of the Neural Pub/Sub system, demonstrating both progress and the remaining challenges in making this innovative paradigm a reality in real-world applications.

## VI. CONCLUSION

This paper introduced the Neural Pub/Sub paradigm, a novel approach to orchestrating AI models and AI-powered applications across the computing continuum. Through the utilization of mapping and funnel patterns, the paradigm offers a flexible framework for distributed data processing, enabling dynamic placement and parallel processing. The ongoing development of a distributed testbed across two university sites further demonstrates the feasibility of this novel paradigm. Open challenges remain in system design, model optimization, and real-world implementation, but the groundwork laid here offers promising pathways for further research and practical application.

## REFERENCES

[1] H. Kokkonen, L. Lovén, N. H. Motlagh, A. Kumar, J. Partala, T. Nguyen, V. C. Pujol, P. Kostakos, T. Leppänen, A. González-Gil et al., "Autonomy and intelligence in the computing continuum: Challenges, enablers, and future directions for orchestration," *arXiv preprint arXiv:2205.01423*, 2022.

[2] T. Taleb, C. Benzaïd, M. B. Lopez, K. Mikhaylov, S. Tarkoma, P. Kostakos, N. H. Mahmood, P. Pirinen, M. Matinmikko-Blue, M. Latva-Aho et al., "6g system architecture: A service of services vision," *ITU journal on future and evolving technologies*, vol. 3, no. 3, pp. 710–743, 2022.

[3] X. Liu, J. Yu, Y. Liu, Y. Gao, T. Mahmoodi, S. Lambotharan, and D. H. Tsang, "Distributed intelligence in wireless networks," *arXiv preprint arXiv:2208.00545*, 2022.

[4] D. Xu, T. Li, Y. Li, X. Su, S. Tarkoma, T. Jiang, J. Crowcroft, and P. Hui, "Edge intelligence: Empowering intelligence to the edge of network," *Proceedings of the IEEE*, vol. 109, no. 11, pp. 1778–1837, 2021.

[5] Y. Cheng, D. Wang, P. Zhou, and T. Zhang, "A survey of model compression and acceleration for deep neural networks," *arXiv preprint arXiv:1710.09282*, 2017.

[6] E. Villar-Rodriguez, M. A. Pérez, A. I. Torre-Bastida, C. R. Senderos, and J. López-de Armentia, "Edge intelligence secure frameworks: Current state and future challenges," *Computers & Security*, p. 103278, 2023.

[7] J. Konečný, H. B. McMahan, F. X. Yu, P. Richtárik, A. T. Suresh, and D. Bacon, "Federated learning: Strategies for improving communication efficiency," *arXiv preprint arXiv:1610.05492*, 2016.

[8] S. Yi, C. Li, and Q. Li, "A survey of fog computing: concepts, applications and issues," in *Proceedings of the 2015 workshop on mobile big data*, 2015, pp. 37–42.

[9] B. Costa, J. Bachiega Jr, L. R. de Carvalho, and A. P. Araujo, "Orchestration in fog computing: A comprehensive survey," *ACM Computing Surveys (CSUR)*, vol. 55, no. 2, pp. 1–34, 2022.

[10] S. Tarkoma, *Publish/subscribe systems: design and principles*. John Wiley & Sons, 2012.

[11] A. Carzaniga, D. S. Rosenblum, and A. L. Wolf, "Design and evaluation of a wide-area event notification service," *ACM Transactions on Computer Systems (TOCS)*, vol. 19, no. 3, pp. 332–383, 2001.

[12] A. Sheth, C. Henson, and S. S. Sahoo, "Semantic sensor web," *IEEE Internet computing*, vol. 12, no. 4, pp. 78–83, 2008.

[13] J. Tirana, C. Pappas, D. Chatzopoulos, S. Lalis, and M. Vavalis, "The role of compute nodes in privacy-aware decentralized ai," in *Proceedings of the 6th International Workshop on Embedded and Mobile Deep Learning*, 2022, pp. 19–24.

[14] Y. Shen, J. Shao, X. Zhang, Z. Lin, H. Pan, D. Li, J. Zhang, and K. B. Letaief, "Large language models empowered autonomous edge ai for connected intelligence," *arXiv preprint arXiv:2307.02779*, 2023.